\documentstyle[aaspp4]{article}
\def\kms {km~s$^{-1}$}

\begin{document}
\title{Does Infall End Before the Class I Stage?}
\author{Erik M. Gregersen\altaffilmark{1}}
\affil{Department of Physics and Astronomy, McMaster
University, Hamilton, ON L8S 4M1, Canada and Department of Astronomy, California Institute of Technology, MS 105-24,
Pasadena, CA 91125}
\author{Neal J. Evans II\altaffilmark{2}}
\affil{Department of Astronomy, The University of Texas at Austin,
       Austin, TX 78712--1083}
\author{Diego Mardones\altaffilmark{3}} 
\affil{Departamento de Astronomia, Universidad de Chile, Casilla 36-D, Santiago, Chile}
\author{Philip C. Myers\altaffilmark{4}}
\affil{Harvard-Smithsonian Center for Astrophysics, 60 Garden Street, Cambridge, MA 02138}
\altaffiltext{1}{Electronic mail: gregerse@physun.physics.mcmaster.ca}
\altaffiltext{2}{Electronic mail: nje@astro.as.utexas.edu}
\altaffiltext{3}{Electronic mail: mardones@das.uchile.cl}
\altaffiltext{4}{Electronic mail: pmyers@cfa.harvard.edu}
\centerline{\footnotesize {\LaTeX}ed at \number\time\ min., \today}
\begin{abstract}

We have observed HCO$^{+}$ $J=3-2$ toward 16 Class I sources and 18 Class 0 
sources, many of which were selected from Mardones et al. (1997). Eight sources 
have profiles significantly skewed to the blue relative to optically thin 
lines.  We suggest six sources as new infall candidates.  We find an equal 
``blue excess" among Class 0 and Class I sources after combining this sample 
with that of Gregersen et al.\ (1997).  We used a Monte Carlo code to simulate 
the temporal evolution of line profiles of optically thick lines of HCO$^{+}$, 
CS and H$_{2}$CO in a collapsing cloud and found that HCO$^{+}$ had the 
strongest asymmetry at late times.  If a blue-peaked line profile implies 
infall, then the dividing line between the two classes does not trace the end 
of the infall stage.    
\end{abstract}


\section {INTRODUCTION}

For many years, infalling protostars were sought, but most claims of collapse 
were challenged.  Through the discovery of 
a population of cold ($T_D$ $\sim$ 30 
K), heavily enshrouded ($A_V$ $\sim$ 1000) objects by IRAS and groundbased
telescopes, a number of protostellar candidate objects became available for
study. 
Subsequent observations of these Class 0 sources 
(Andr\'e et al.\ 1993) revealed spectral-line profiles 
indicative of infall in 13 objects (Walker et al.\ 1986; Zhou et al.\ 
1993; Zhou et al. 1994; Moriarty-Schieven et al.\ 1995; Hurt et al.\ 1996; 
Myers et al.\ 1996; Ward-Thompson et al.\ 1996; Gregersen et al.\ 1997; 
Lehtinen 1997; Mardones et al.\ 1997).  If these infall candidates actually 
represent infall, we can begin 
to study the evolution of infall by comparing line profiles of sources believed 
to be in different evolutionary states. 

Mardones et al.\ observed 23 Class 0 sources ($T_{bol}$ $<$ 70 K) and 24 Class 
I sources (70 $\le$ $T_{bol}$ $\le$ 200 K) in the optically thick H$_{2}$CO 
$2_{12}-1_{11}$ and CS $J=2-1$ lines and the optically thin N$_{2}$H$^{+}$ 
$JF_{1}F_{2}=101-012$ line.  $T_{bol}$, the temperature of a blackbody with the 
same mean frequency as the observed spectral energy distribution, increases 
with age from the highly embedded Class 0 sources to the pre-main sequence 
Class III sources ($T_{bol}$ $>$ 2800 K) (Myers and Ladd 1993).  The upper 
boundary of the Class I category is 650 K but Mardones et al.\ chose the most 
deeply embedded Class I sources to study the earliest stages of accretion.  
Since protostellar collapse is expected to produce double-peaked profiles in 
optically thick lines with the blue peak stronger than the red peak or lines 
that are blue-skewed relative to the velocity of an optically thin line, 
Mardones et al.\ compared the percentage of sources in each Class that 
displayed such asymmetry and found that the ``blue excess" (the number of 
sources with significant blue asymmetry  minus the number of sources with 
significant red asymmetry divided by the total number of sources) was very 
different for Class 0 and Class I sources: for H$_{2}$CO, 0.39 for Class 0 
sources and 0.04 for Class I sources; and for CS, 0.53 for Class 0 sources and 
0.00 for Class I sources.  Park et al.\ (1999) have done a survey of Class 0 
and I sources in HCN $J=1-0$ and observed similar results.  Using the 
significance criterion of Mardones et al., the blue excess is 0.27 for Class 0 
sources and 0.00 for Class I sources.

We decided to observe the Class I sources studied in Mardones et al.\ in 
HCO$^{+}$ $J=3-2$ for two reasons.  First, the data obtained would be more 
easily compared with the HCO$^{+}$ Class 0 survey of Gregersen et al.\ (1997).  
Second, the HCO$^{+}$ $J=3-2$ line is usually stronger and more opaque than the 
other lines; consequently, it may be more sensitive to infall at later stages 
when much of the cloud material has accreted but infall is not complete.

The sources we observed are listed in Table 1.  Our sample includes 4 
Class 0 sources Mardones et al.\ observed that were not observed by Gregersen 
et al.\ as well as 10 sources not observed by Mardones et al.\ but which have 
been identified by others as either Class 0 or early Class I.

\section {OBSERVATIONS AND RESULTS}

We observed 16 Class I and 18 Class 0 sources in the HCO$^{+}$ $J=3-2$ line 
with the 10.4-m telescope of the Caltech Submillimeter Observatory 
(CSO)\footnote{The CSO is operated by the California Institute of Technology 
under funding from the National Science Foundation, contract AST 90--15755.} at 
Mauna Kea, Hawaii in December 1995, September 1996, April 1997, June 1997, 
December 1997 and July 1998.  The sources are listed in Table 1 with their 
celestial coordinates, distances, $T_{bol}$ and the off position used for 
position switching.  We used an SIS receiver (Kooi 1992).  The backend was an 
acousto-optic spectrometer with 1024 channels and a bandwidth of 49.5 MHz.  The 
frequency resolution was about 2 channels, which is 0.11 \kms\ at 267 GHz, 
except for the December 1995 observations when the resolution was near 3 
channels or 0.16 \kms\ at 267 GHz.  Chopper-wheel calibration was used to 
obtain the antenna temperature, $T_A^*$.  The lines we observed are listed in 
Table 2 with their frequencies, the velocity resolution, the beamsize and the 
main beam efficiency, $\eta_{mb}$.  The main beam efficiencies were calculated 
using planets as calibration sources.  Data from separate runs were resampled 
to the resolution of the run with the worst frequency resolution before 
averaging.  A first order baseline was removed before spectra were averaged.  

Line properties are listed in Table 3.  $T_A^*$ is the peak temperature in the 
line profile.  For single-peaked lines, $V_{LSR}$, the line centroid, and 
$\Delta$V, the line width, were found by fitting a single Gaussian to the line 
profile.  For lines that have two blended peaks, we list two values of $T_A^*$ 
and $V_{LSR}$, one for each peak, and one value for the line width, which is 
the width across the spectrum at the temperature where the weaker peak falls to 
half power.

We observed 34 sources in the HCO$^{+}$ $J=3-2$ line.  Nine of these sources
showed blue asymmetry (Figure 1) and six showed red asymmetry (Figure 2) 
as determined by visual inspection of the line profiles.  Nineteen sources
showed no significant asymmetry (Figures 3 and 4).  We observed eight 
sources in the H$^{13}$CO$^{+}$ $J=3-2$ line, five of which showed blue
asymmetry in the HCO$^{+}$ $J=3-2$ line. 

\section {Individual Sources}

\paragraph {IRAS03235+3004}

Ladd et al.\ (1993) observed this source at H and K.  Mardones et al.\ (1997)
observed a red-skewed line profile in both CS and H$_{2}$CO.  The HCO$^{+}$ 
$J=3-2$ line is double-peaked with the blue peak stronger (Figure 1).  Both the 
H$_{2}$CO and CS lines of Mardones et al.\ peak at the HCO$^{+}$ red peak while 
the N$_{2}$H$^{+}$ line peaks slightly to the blue of the HCO$^{+}$ dip.  
However, there is significant H$_{2}$CO and CS emission at the velocity of the 
blue peak.

\paragraph {L1455}

Frerking and Langer (1982) first detected a CO outflow in this source.  
Submillimeter observations revealed a small dense core surrounding the
exciting source (Davidson and Jaffe 1984).  Goldsmith et al.\ (1984) observed 
two well-collimated outflows.  Mardones et al.\ observed a symmetric H$_{2}$CO 
line.  We observe a slightly blue-skewed line in HCO$^{+}$ $J=3-2$ which peaks 
0.5 \kms\ to the blue of the H$_{2}$CO line (Figure 3).

\paragraph {IRAS03256+3055}

This source is located near the NGC1333 complex.  Mardones et al.\ observed a 
red-peaked line profile in both lines with the red peak at the N$_{2}$H$^{+}$ 
velocity.  The HCO$^{+}$ profile we see is symmetric (Figure 3).

\paragraph {SSV13}

The Herbig-Haro objects HH 7-11 are excited by this source.  Warin et al.\
(1996) have suggested that this source is triggering star formation in NGC 
1333.  Outflow wings are quite prominent in the CS, H$_{2}$CO (Mardones et 
al.) and HCO$^{+}$ profiles (Figure 4), but no significant skewness exists in
the central core of the line.

\paragraph {NGC 1333 IRAS 2}

Sandell et al.\ (1994) and Hodapp and Ladd (1995) observed two outflows, 
suggesting that this source may be a protobinary.  Ward-Thompson et al.\ (1996) 
modeled the HCO$^{+}$ and H$^{13}$CO$^{+}$ $J=4-3$ spectra as infall.  Both the 
CS and H$_{2}$CO lines observed by Mardones et al.\ are skewed to the red.  The 
HCO$^{+}$ $J=3-2$ has the same self-absorption dip and blue-skewed profile as 
the $J=4-3$ spectra (Figure 1).  

\paragraph {IRAS03282+3035}

Bachiller er al.\ (1991) observed a high velocity, well-collimated outflow in
this source.  Bachiller and Gomez-Gonzales (1992) identified this as an
``extreme Class I" object, a category Andr\'e et al.\ (1993) later called 
Class 0.  The HCO$^{+}$ $J=3-2$ has a slight blue asymmetry (Figure 3)
but it is insufficient to qualify this source as a collapse candidate.

\paragraph {HH211}

McCaughrean et al.\ (1994) discovered a molecular hydrogen jet with a dynamical
age of less than 1000 years excited by a source detected only at wavelengths
greater than 350 $\mu$m.  Like the HCO$^{+}$ $J=3-2$ spectra, the CS and 
H$_{2}$CO lines have no significant asymmetry (Mardones et al.\ 1997) (Figure 4).

\paragraph {IRAS04166+2706}

Mardones et al.\ observed a symmetric line in H$_{2}$CO and a double-peaked 
line in CS with the stronger peak at the central velocity.  The HCO$^{+}$ 
$J=3-2$ line is nearly symmetric with a weak dip on the red-shifted side 
(Figure 1).  The weaker of the two CS peaks lies within the blue HCO$^{+}$ line wing.

\paragraph {IRAS04169+2702}

Ohashi et al.\ (1997) detected evidence for infall in a 2200 $\times$ 1100 AU 
envelope around this source using channel maps from interferometric 
observations of C$^{18}$O $J=1-0$.  Mardones et al.\ saw outflow wings 
in CS.  Our HCO$^{+}$ $J=3-2$ data shows a symmetric profile (Figure 3), 
which encompasses the range of infall velocities seen by Ohashi et al., 
at the rest velocity of the source.

\paragraph {L1551 IRS5}

The bipolar outflow was first seen in this source (Snell et al.\ 1980).
Butner et al.\ (1991) found a density gradient consistent with the collapse 
model of Adams et al.\ (1987).  Submillimeter interferometry revealed an 80 AU 
disk (Lay et al.\ 1994).  Looney et al.\ (1997) interpreted their 2.7 mm 
interferometer observations as evidence that this source is a protobinary with 
each source separated by 50 AU (too small to be resolved by Lay et al.) and 
having a disk with a radius $<$ 25 AU. Ohashi et al.\ (1996) observed in 
$^{13}$CO $J=2-1$ a much larger central condensation, 1200 by 670 AU, which had 
infalling motions.  The H$_{2}$CO and CS lines (Mardones et al.), as well as 
our HCO$^{+}$ $J=3-2$ line, are symmetric with outflow wings (Figure 3).

\paragraph {L1535}

The H$_{2}$CO and CS observations of Mardones et al.\ show a symmetric 
line profile, as do our HCO$^{+}$ $J=3-2$ data (Figure 3).

\paragraph {TMC-1A}

Wu et al.\ (1992) discovered a CO outflow in this source.  CO $J=1-0$ 
interferometer observations suggested the existence of a 2500 AU flattened 
structure (Tamura et al.\ 1996).  Chandler et al.\ (1996) noted the similarity 
of the outflow to those of the Class 0 sources L1448-C and B335.  Ohashi et 
al.\ (1997) measured a velocity gradient that would be expected from a rotating 
disk perpendicular to the outflow over a 580 AU radius.  Mardones et al.\ 
observed symmetric lines in H$_{2}$CO and CS.  The HCO$^{+}$ $J=3-2$ line is 
double peaked with a stronger blue peak, but the N$_{2}$H$^{+}$ line of 
Mardones et al.\ (presumed to be optically thin) is near the red peak rather 
than between the two (Figure 1).  The red peak of the HCO$^{+}$ $J=3-2$ line is 
coincident with the peak of the CS line.  There is no CS emission at the 
velocity of the blue peak, suggesting that the blue peak might be outflow 
emission.

\paragraph {L1634}

A powerful molecular hydrogen jet is present in this source (Hodapp and Ladd 
1995; Davis et al.\ 1997).  The HCO$^{+}$ spectra has the blue peak slightly 
stronger than the red peak (Figure 1).

\paragraph {MMS1}

This source and the following five were discovered by Chini et al.\ (1997), in 
the Orion high-mass star-forming region.  Most of the other sources observed in
this paper occur in low-mass star-forming regions.  They were identified as 
Class 0 objects from their strong millimeter continuum emission.  The HCO$^{+}$ 
$J=3-2$ has a double-peaked profile with the red peak stronger (Figure 2).

\paragraph {MMS4}

The red peak is slightly stronger than the blue peak in the HCO$^{+}$ $J=3-2$
spectra (Figure 2).

\paragraph {MMS6}

Little noticeable asymmetry is seen in the HCO$^{+}$ $J=3-2$ spectra (Figure 3).

\paragraph {MMS9}

The HCO$^{+}$ $J=3-2$ line is symmetric (Figure 4).

\paragraph {MMS7}

The HCO$^{+}$ $J=3-2$ line has a slight red shoulder (Figure 4).

\paragraph {MMS8}

This source has a  highly collimated CO outflow (Chini et al.\ 1997).
No asymmetry is observed in the HCO$^{+}$ $J=3-2$ spectra (Figure 3).

\paragraph {NGC2264G}

Margulis and Lada (1986) discovered the molecular outflow which was later seen
to be highly collimated and energetic (Margulis et al.\ 1988).  Margulis
et al.\ (1990) observed six near-infrared sources in this object, but Gomez
et al.\ (1994) discovered the true driving source of the outflow, which
was subsequently confirmed as a Class 0 object by Ward-Thompson et al.\
(1995).  The HCO$^{+}$ $J=3-2$ spectrum has a slight red wing (Figure 3). 

\paragraph {B228}

Heyer and Graham (1989) found evidence for a stellar wind from observations
of extended [SiII] emission.  Mardones et al.\ observed a slightly blue-skewed 
H$_{2}$CO line.  The HCO$^{+}$ $J=3-2$ line has a double peaked profile with a 
stronger blue peak like that predicted by collapse models (Figure 1).  The 
H$^{13}$CO$^{+}$ $J=3-2$ line is coincident with the red peak, but the 
N$_{2}$H$^{+}$ line peaks in the self-absorption dip of the HCO$^{+}$ line, 
suggesting that the H$^{13}$CO$^{+}$ line may be somewhat optically thick.

\paragraph {YLW16}

The H$_{2}$CO has a prominent blue outflow shoulder (Mardones et al.).  A 
similar feature appears in the HCO$^{+}$ $J=3-2$ line (Figure 4).

\paragraph {L146}

Clemens et al.\ (1998) identified this as a Class 0 or I source, an 
identification that corresponds well with its $T_{bol}$ of 74 K.  Mardones
et al.\ observed a double-peaked line in CS and a blue-shouldered line in
H$_{2}$CO.  The HCO$^{+}$ $J=3-2$ line is symmetric with a hint of a blue 
shoulder and agrees in velocity with the red peak in the CS spectrum (Figure 
3).

\paragraph {S68N}

McMullin et al.\ (1994) first identified this core from CS and CH$_{3}$OH maps.
Hurt and Barsony (1996) established this as a Class 0 source.   Wolf-Chase et 
al.\ (1998) found an outflow flux consistent with those of other Class 0 
sources.  The CS and H$_{2}$CO lines show blue asymmetry with a prominent
outflow shoulder (Mardones et al.).  The HCO$^{+}$ $J=3-2$ line has strong
self-absorption with the red peak stronger than the blue peak.  The
red peak has the same velocity as the Mardones et al.\ spectra, while the 
blue peak has the same velocity as that of the CS profile.  The H$^{13}$CO$^{+}$
line occurs at the same velocity as the self-absorption dip (Figure 2)

\paragraph {SMM6}

This source is better known as SVS20 (Strom et al.\ 1976).  The HCO$^{+}$
spectra is symmetric (Figure 4).

\paragraph {HH108}

The Herbig-Haro objects, HH108 and HH109, are produced by a 9.5 L$_{\sun}$
IRAS source (Reipurth and Eiroa 1992).  Reipurth et al.\ (1993) detected
strong 1.3 mm emission.  Both the CS $J=2-1$ and H$_{2}$CO $J=2_{12}-1_{11}$ 
lines have peaks shifted to the red of the N$_{2}$H$^{+}$ velocity (Mardones
et al.).  Our HCO$^{+}$ line profile looks similar to that of the H$_{2}$CO 
$J=2_{12}-1_{11}$ line but with a slightly stronger blue peak (Figure 2).

\paragraph {CrA IRAS32}

Both the H$_{2}$CO $J=2_{12}-1_{11}$ and CS $J=2-1$ lines are symmetric
(Mardones et al.).  The HCO$^{+}$ $J=3-2$ line has a double-peaked profile 
with a strong red peak and a dip near the velocity of the N$_{2}$H$^{+}$ line 
(Figure 2).  The H$_{2}$CO and CS lines also peak in the HCO$^{+}$ dip.

\paragraph {L673A}

This core appears slightly elongated in continuum emission and is about 0.2 pc 
away from another core (Ladd et al.\ 1991).  Like the H$_{2}$CO and CS spectra 
(Mardones et al.), the HCO$^{+}$ $J=3-2$ line is symmetric (Figure 4).

\paragraph {L1152}

The H$_{2}$CO $J=2_{12}-1_{11}$ and CS $J=2-1$ lines are symmetric (Mardones 
et al.), as is the HCO$^{+}$ $J=3-2$ line (Figure 4).

\paragraph {L1172}

Submillimeter continuum maps of Ladd et al.\ (1991) showed extended emission
at 100 and 160 $\mu$m.  The CS $J=2-1$ has a narrow self-absorption dip 
with the blue peak slightly stronger than the red peak (Mardones
et al.).  The HCO$^{+}$ spectrum (Figure 2) is slightly red-peaked and the two 
peaks agree in velocity with those in the CS spectrum (Mardones et al.).
The N$_{2}$H$^{+}$ line has its peak in the HCO$^{+}$ self-absorption dip.

\paragraph {L1251A}

This source is located near the edge of its cloud and may be older than
L1251B (Sato and Fukui 1989).  The peak of the H$_{2}$CO line is slightly
blue-shifted.  The HCO$^{+}$ line is symmetric with prominent outflow
wings (Figure 4).

\paragraph {L1251B}

Sato and Fukui (1989) discovered a molecular outflow in this source. Myers et 
al.\ (1996) matched a simple infall model to observed CS $J=2-1$ and 
N$_{2}$H$^{+}$ $JF_{1}F=101-012$ lines.  The H$_{2}$CO $J=2_{12}-1_{11}$ line
observed by Mardones et al.\ (1997) displays prominent self-absorption.  The 
HCO$^{+}$ line profile is very strong with a prominent self-absorption dip and 
has the same velocity structure as the CS spectrum.  The H$^{13}$CO$^{+}$ 
$J=3-2$ line lies at the velocity of the dip, indicating that this may be a 
good candidate for protostellar collapse (Figure 1).

\paragraph {IRAS23011}

Lefloch et al.\ (1996) identified this as a Class 0 source based on its 
millimeter continuum emission and its highly energetic outflow.
Ladd and Hodapp (1997) observed a double outflow and measured a bolometric
temperature of 61 K.  The HCO$^{+}$ $J=3-2$ has collapse asymmetry with
the H$^{13}$CO$^{+}$ $J=3-2$ lying between the two peaks (Figure 1).

\paragraph {CB244}

Launhardt et al.\ (1997) identified this as a Class 0 object from its
spectral energy distribution.  The CS spectrum is symmetric, but the line is 
blue-shifted.  The H$_{2}$CO $J=2_{12}-1_{11}$ line is blue-shifted with a 
prominent outflow wing. The HCO$^{+}$ $J=3-2$ shows a blue-peaked line profile 
suggestive of collapse (Figure 1).  The N$_{2}$H$^{+}$ line observed by 
Mardones et al.\ is at the velocity of the self-absorption dip.  The red peak 
in the HCO$^{+}$ $J=3-2$ spectrum is at the same velocity as the H$_{2}$CO 
peak.  The blue HCO$^{+}$ peak is at the same velocity as the CS line.

\section {DISCUSSION}

\subsection{Line Profile Asymmetry}

Models of a collapsing source with appropriate excitation requirements, a 
temperature and density gradient increasing toward the interior, predict that 
optically thick lines should show a double-peaked line profile with the 
blue-shifted peak stronger than the red-shifted peak and a self-absorption dip 
caused by the envelope at the rest velocity of the source (Leung and Brown 
1977, Walker et al.\ 1986, Zhou 1992).  The 
ratio of the peak temperatures of the blue and red peaks is a natural 
way to quantify this asymmetry, but this method requires lines that are 
double-peaked.

Since many of our line profiles are asymmetric without possessing two peaks, we 
also quantified the asymmetry using the asymmetry parameter defined in Mardones 
et al.\ as 
$$
\delta V = 
{(V_{thick} - V_{thin})\over{\Delta v_{thin}}}
$$
where $V_{thick}$ is the velocity of the peak of the optically thick HCO$^{+}$ 
$J=3-2$ line , $V_{thin}$ is that of the optically thin line as determined by a 
Gaussian fit,  and $\Delta v_{thin}$ is the line width of the optically thin 
line.  The optically thin line we used was H$^{13}$CO$^{+}$ $J=3-2$ or, where 
that was not available, the N$_{2}$H$^{+}$ $JF_{1}F=101-012$ line ($\nu$ = 
93.176265 GHz, Caselli et al.\ 1995) observed by Mardones et al.  In Figure 5, 
we plot a histogram of the velocity of the H$^{13}$CO$^{+}$ line minus
that of the N$_{2}$H$^{+}$ line.  The mean velocity difference between
the two lines is 0.15 $\pm$ 0.20 \kms.  
Following Mardones et al., we use $\delta V$ of $\pm$0.25 as the threshold to 
be counted as either blue (if negative) or red (if positive).  Also, we follow 
Mardones et al.\ in not considering an asymmetry as significant if the 
difference in strength between the blue and red peaks is less than twice the 
rms noise.  For example, in L1634, the blue peak is only  0.11 K stronger than 
the red peak while the rms noise is 0.10 K; we do not consider this source to 
be an infall candidate. The results for $\delta V$ are listed in Table 4 for 
the objects observed in this paper. 

Among these sources, there are more blue than red objects.  We can quantify 
this by using the blue excess, which is defined by Mardones et al.\ as
$$
Blue\ Excess = {N_{blue} - N_{red}\over{N_{total}}}.
$$ 
If the blue line asymmetries expected from collapse are dominant in our 
sample, this will be a positive number.  
For the sources observed in this paper, the blue excess is 0.28.

\subsection{Evolutionary Trends of the Line Profile}

The sources observed in this paper can be combined with the 20 Class 0 and
3 Class I sources observed in Gregersen et al.\ (1997) to form a sample
stretching from $T_{bol}$ of 30 to 170 K, from the Class 0 stage well into
the Class I stage.  As in the previous section, we use for the optically thin 
line the H$^{13}$CO$^{+}$ $J=3-2$ line when available and the N$_{2}$H$^{+}$ 
$JF_{1}F=101-012$ line when it is not.  The results for $\delta V$ 
are plotted versus bolometric temperature in Figure 6.

{\it For $\delta V$, there is no clear trend with bolometric temperature.} We 
find a significant blue excess among both Class 0 and I sources.  There is no 
dividing line past which the blue excess drops, a result markedly different 
from that of Mardones et al.\ who saw a disappearance of blue-skewed line 
profiles by the Class I stage.  In fact, the blue peaked TMC-1A has the highest 
$T_{bol}$, 170 K, of any source in our sample.  The results for $\delta V$ 
for each class as well as for the combined sample are listed in Table 5. 

To study the difference between lines, we plot $\delta V$ as measured by the 
HCO$^{+}$ against those measured by the H$_{2}$CO and CS lines (Figure 7).  The 
sources that appear in our blue excess and not those of Mardones et al.\ are in 
the box bounded by $\delta V$ (H$_{2}$CO or CS) $> -0.25$ and 
$\delta V$(HCO$^{+}$) $< -0.25$.  For example, TMC-1A and IRAS03235, sources 
with pronounced blue $\delta V$ in HCO$^{+}$, show little asymmetry in CS and 
H$_{2}$CO.  However, most of the differences between the tracers involve 
sources with no significant asymmetry in CS or H$_{2}$CO, but a significant one
in HCO$^{+}$. Only a few sources have significant disagreements (upper left
and lower right boxes outside the dashed lines in Figure 7). This pattern
would be expected if the HCO$^{+}$ were more sensitive to infall in envelopes
of lower opacity.
 
There are some sources that show markedly opposite asymmetries in CS or 
H$_{2}$CO as opposed to HCO$^{+}$.  In the upper left corner of both plots, 
L1551-NE, a Class 0 source within an outflow lobe of L1551 IRS5, has a 
complicated HCO$^{+}$ $J=3-2$ triple-peaked spectra (Gregersen et al.) with a 
very strong blue peak.  The H$_{2}$CO and CS spectra are also triple-peaked, 
but the peaks are not so prominent that Mardones et al.\ were unable to fit a 
Gaussian to the line. We do not consider this source to be an infall candidate.
Fitting a Gaussian in this case will give a redder velocity and thus a positive 
$\delta V$.  In the lower right corner of the H$_{2}$CO-HCO$^{+}$ plot is L483, 
which has an H$_{2}$CO spectra with similar velocity structure but is 
blue-peaked in H$_{2}$CO and red-peaked in HCO$^{+}$.   

Mardones et al.\ found good correlation between the CS and H$_{2}$CO
$\delta V$ (the two measures agree in 28 out of 38 instances) and Park et al.\
(1999) also found a good correlation between CS and the HCN $J=1-0$ $F=2-1$ 
line ($\nu$ = 88.631847 GHz) and a fair correlation between H$_{2}$CO and HCN.
We consider agreement as when both tracers are significantly blue or red 
or symmetric, disagreement as when one is significantly blue and the other 
significantly red and neutrality when one is significantly blue or red
and the other is symmetric.  CS and HCO$^{+}$ agree 13 times, are neutral
15 times and disagree 6 times.  For H$_{2}$CO, there is agreement 16 times,
neutrality 16 times and disagreement 5 times.  For HCN, the asymmetries
agree 7 times, are neutral 6 times and disagree 4 times.

\subsection{An Evolutionary Model}

Unlike the CS and H$_{2}$CO observations of Mardones et al., our HCO$^{+}$
observations of Class I sources do not show a radical change in the excess of
sources with blue asymmetry at the Class 0-Class I boundary.  Why do the
HCO$^{+}$ results differ from those of HCN, CS and H$_{2}$CO? Perhaps infall 
asymmetry disappears at later times in HCO$^{+}$ than in HCN, CS and H$_{2}$CO.

For models in which the infall velocity increases closer to the forming star,
each line of sight through the infall region intersects two points on
the locus of constant radial velocity. 
To display infall asymmetry, a spectral line must be subthermally excited
and fairly opaque at the foreground point of intersection. Different lines
will be more suitable at different points in the evolution, as the density 
drops. Among the lines in Table 6, the HCO$^{+}$ and H$_2$CO lines require the 
highest densities to excite (see Table 1 in Evans 1999) and HCO$^{+}$ has a 
higher opacity for typical abundances. Therefore, it is possible that the 
HCO$^{+}$ $J=3-2$ line traces late stages of infall, when densities and 
opacities are lower, than the other lines.

To test this possibility and to study how the infall signature changes with 
time, one of the HCO$^{+}$ evolutionary models presented in Gregersen et al.\ 
has been extended to later times and collapse models have done for CS and 
H$_{2}$CO at those same times.  The lines modeled are HCO$^{+}$ $J=3-2$, CS 
$J=2-1$ and H$_{2}$CO $J=2_{12}-1_{11}$, the lines observed by Gregersen et al. 
and Mardones et al., respectively.  (The CS line has been modeled for early times 
by Zhou (1992) and Walker et al.\ (1994).)  Protostellar collapse was 
simulated in a 
cloud with a radius of 0.2 pc using the velocity and density fields of the Shu 
(1977) collapse model and a temperature distribution scaled upward from that of 
B335 (Zhou et al.\ 1990) to a luminosity of 6.5 L$_{\sun}$, the average 
luminosity of the sources observed in Gregersen et al.  Six models were run for 
infall radii of 0.005, 0.03, 0.06, 0.10, 0.13 and 0.16 pc, corresponding to 
infall times of 2.3 $\times$ 10$^{4}$ yr for the earliest model to 7.5 $\times$ 
10$^{5}$ yr for the latest model.  The cloud had 30 shells, 15 inside the 
infall radius and 15 outside.  The model produces the velocity, density, 
kinetic temperature, turbulent width and molecular abundance for each shell.  
The same abundance (6 $\times$ 10$^{-9}$)  was used for each molecule so these 
models would be the simplest possible comparisons of molecular tracers.   

The output of the collapsing cloud simulation was used as input for a Monte 
Carlo code (Choi et al.\ 1995), which produces molecular populations in each 
shell.  The output of the Monte Carlo code was used as input for a program that 
convolves a Gaussian beam with the emission from the cloud so we can simulate
our HCO$^{+}$ CSO observations and the H$_{2}$CO IRAM and CS Haystack
observations of Mardones et al.  The distance to the model cloud was 310 pc,
the average distance to the observed sources.
The resulting line profiles are plotted in Figure 8.

The HCO$^{+}$ line profiles are the strongest and, at late times, the most 
asymmetric.  The dashed horizontal line across each panel is the blue-red 
ratio.  For CS and H$_{2}$CO, the blue-red ratio reaches a peak value and 
levels off, but the HCO$^{+}$ blue-red value keeps increasing.  For all lines, 
the velocity of the blue-shifted peak becomes more negative with time from 
about $-0.22$ \kms\ to about $-0.33$ \kms. 
The profiles in Figure 8 support the conclusion that HCO$^{+}$ $J=3-2$
reveals infall more readily than the CS and H$_2$CO lines. However,
the model line profiles for those molecules predict a detectable signature as well.
One possibility for obscuring the signature in other lines 
is that HCO$^{+}$ remains in the gas phase while the other
species freeze out (Rawlings et al. 1992,  Park et al. 1999). 

One can question the relevance of our simulation, because none of the observed 
sources can be completely explained by the simple inside-out collapse model.  
For example, the simulated line profiles often have less extreme blue-red 
ratios than most of our infall candidates and the dip is much more extreme than
in the observations.  All of these sources are either turbulent, aspherical,
or have very energetic bipolar outflows.  Outflow is a large problem for 
HCO$^{+}$ $J=3-2$ because emission from outflows is prominent in this line.
Also, recent studies of pre-protostellar cores (e.g. Tafalla et al.\ 1999),
cores seemingly in the earliest stages of collapse, have challenged the 
inside-out collapse model.

On the whole, the HCO$^{+}$ $J=3-2$ line has both advantages and disadvantages
in searching for infall. In the following section, we consider the information
from all lines in assessing the best infall candidates.

\subsection{New Infall Candidates?}

Among the 34 sources observed here, we see 8 sources with the correct 
asymmetry for infall (Figure 1). (L1634 is eliminated from consideration for
its weak asymmetry.)  Three of these sources are Class 0, four
are Class I and two are probably Class I.  However, a blue-peaked optically 
thick line alone is not enough for a definite claim of infall.  

For such a claim,  the optically thin line must peak in the self-absorption dip 
of the optically thick line.  We have observed the H$^{13}$CO$^{+}$ $J=3-2$ 
line in 4 of our sources.  In L1251B, NGC 1333 IRAS 2 and IRAS 23011, this 
condition is met.  For sources where we did not observe the H$^{13}$CO$^{+}$ 
$J=3-2$ line, we used the N$_{2}$H$^{+}$ $J=1-0$ line of Mardones et al.\ to 
provide a rest velocity.  Among the sources with blue peaks, N$_{2}$H$^{+}$ 
peaks in the dip of CB244, TMC-1A, IRAS 03235 and IRAS 04166, so we consider 
these to be good infall candidates.  In B228, the N$_{2}$H$^{+}$ line is at the 
same velocity as the HCO$^{+}$ self-absorption dip, so we believe the 
H$^{13}$CO$^{+}$ $J=3-2$ line may be optically thick in this source and we 
consider it to be a possible infall candidate.   
All the candidates that could have passed this test did so.

Since these sources have been observed in three optically thick lines, CS, 
H$_{2}$CO and HCO$^{+}$, with roughly comparable beamsizes, we can rank the
worthiness of these sources as infall candidates (Table 6).  Sources with a 
blue asymmetry in each of the three lines are the strongest infall
candidates while those with blue asymmetries in two lines are
the next strongest down to sources with one blue asymmetry.  
In each category, sources that display
either blue or no asymmetry in each of the three lines are considered stronger 
candidates than those with red asymmetries.  Sources above B228 are
considered infall candidates.  We include the HCN results of Park 
et al.\ for completeness, but consider them a secondary criteria since their 
beam was 1\arcmin, larger than that in the observations of Gregersen et al.\ 
and Mardones et al.  

Based on these criteria, L1251B is the strongest infall 
candidate of the sources studied here and is followed by CB244.  
Among the 
other six sources with infall asymmetry in HCO$^{+}$, NGC 1333 IRAS 2 has an 
H$_{2}$CO spectrum with a strong red peak and so is ruled out as an infall 
candidate since outflow seems likely as the explanation for the asymmetric 
profiles.

\section {CONCLUSIONS}

We have observed 16 Class I and 18 Class 0 sources in HCO$^{+}$ $J=3-2$.  Nine 
sources have a blue asymmetry and six have a red asymmetry.  Infall asymmetries 
as defined by Mardones et al.\ are still present in Class I sources with 
$T_{bol}$ $<$ 170 K (blue excess = 0.31) to the same extent they are present 
among Class 0 sources with $T_{bol}$ $<$ 70 K (blue excess = 0.31).  If the 
HCO$^{+}$ $J=3-2$ line is more sensitive to infall, as evolutionary models 
suggest, than the CS and H$_{2}$CO lines studied by Mardones et al., then the 
end of the collapse phase still remains to be found.  Among the sources we 
surveyed, we suggest six new infall candidates: CB 244, IRAS 03235, IRAS 04166, 
B228, IRAS23011 and TMC-1A. We confirm the suggestion of Myers et al.\ (1996) 
that L1251B is an infall candidates.   
NGC 1333 IRAS 2 is ruled out as an infall candidate, but has the weakest claim of any 
source studied here since it has blue asymmetry in HCO$^{+}$ and red asymmetry
in H$_{2}$CO.

Although all the sources in this paper are young and protostellar, making an 
unambiguous claim for infall is somewhat difficult. If these sources
are undergoing infall, then surely all
of these sources should have blue asymmetries in every line.  There is the 
optical depth effect previously mentioned where CS and H$_{2}$CO are optically
thinner than HCO$^{+}$.  However, there is also the problem that infall
velocities are roughly the same speed or less than turbulent and outflow motions.
In the infall model previously presented, if such a source was
observed with the resolution of our and Mardones et al.'s observations,
at the radius of our beam, $\sim$ 25\arcsec , for a source at 310 pc,
infall speeds range from .07 \kms\ at the earliest times to 0.60 \kms\
at the latest stages.   In real cores, for example, L1544, a pre-protostellar
core probably at the beginning of the collapse phase, infall speeds are
$\sim$ .1 \kms (Tafalla et al.\ 1999).  
Other caveats include beam size, the particular molecular transition chosen
and the source mass.  Detecting infall is a challenge.

For future work, we should observe CO to determine what velocity ranges
are affected by outflows.  Also, we should observe even later Class I
sources to see if asymmetry is still common.  Work is needed with higher
spatial resolution to separate outflow from infall.  Quantitative infall 
rates also need to be measured to see how the infall rate changes with time. 

\acknowledgments

We would like to thank Daniel Jaffe, Wenbin Li, Kenji Mochizuki and Yancy 
Shirley for their help with observations.  We would also like to thank Daniel 
Jaffe, John Lacy, Charles Lada and John Scalo for their comments on an early 
draft of this paper and the anonymous referee for helpful comments.  
This work was supported by NSF grant AST-9317567.  
D.M. thanks support from FONDECYT grant 1990632 and 
Fundacion Andes grant C-13413/7.

\clearpage
\begin{deluxetable}{llllll}
\footnotesize
\tablewidth{0pt}
\tablecaption{List of Sources}
\tablehead{\colhead{Name} & \colhead{R.A.} & \colhead{Dec.} &
\colhead{Offpos} &\colhead{Distance} & \colhead{T$_{bol}$} \\
& \colhead{(1950.0)} & \colhead{(1950.0)} & \colhead{(\arcsec)} & \colhead{(pc)} & \colhead{(K)}}
\startdata
IRAS03235 & 03:23:33.0 & 30:04:59 & (--1200,0) & 350 & $<$136 \\
L1455 & 03:24:34.9 & 30:02:36 & (--600,0) & 350 & 67 \\
IRAS03256 & 03:25:39.2 & 30:55:20 & (--600,0) & 350 & $<$74 \\
NGC 1333 IRAS 2 & 03:25:49.89 & 31:04:16.3 & (0,--600) & 350 & 52 \\
SSV13 & 03:25:57.9 & 31:05:50 & (--600,0) & 350 & 136 \\
IRAS03282 & 03:28:15.2 & 30:35:14 & (--1200,0) & 350 & 23 \\
HH211 & 03:40:48.7 & 31:54:24 & (--600,0) & 350 & 30 \\
IRAS04166 & 04:16:36.0 & 27:06:00 & (--600,0) & 140 & 91 \\
IRAS04169 & 04:16:54.0 & 27:01:59 & (--600,0) & 140 & 170 \\
L1551 IRS 5 & 04:28:40.2 & 18:01:42 & (--600,0) & 140 & 95 \\
L1535 & 04:32:33.5 & 24:02:15 & (--600,0) & 140 & 61 \\
TMC-1A & 04:36:31.2 & 25:35:56 & (--600,0) & 140 & 170 \\
L1634 & 05:17:21.9 & -05:55:05 & (0,-900) & 415  & 77 \\
MMS1 & 05:32:50.1 & -05:02:13 & (600,0) & 415 & $>$9 \\
MMS4 & 05:32:52.6 & -05:02:46 & (600,0) & 415 & $>$9 \\
MMS6 & 05:32:55.6 & -05:03:25 & (600,0) & 415 & $>$8 \\
MMS9 & 05:32:58.2 & -05:07:35 & (600,0) & 415 & $>$9 \\
MMS7 & 05:32:58.6 & -05:05:46 & (600,0) & 415 & 96 \\
MMS8 & 05:32:58.7 & -05:07:10 & (600,0) & 415 & $>$9 \\
NGC2264G & 06:38:25.76 & 09:58:52.41 & (-600,0) & 800 & 34 \\
B228 & 15:39:50.4 & -33:59:42 & (0,--900) & 150 & 48 \\
YLW 16 & 16:24:26.2 & -24:32:53 & (0,--900) & 125 & 157 \\
L146 & 16:54:27.2 & -16:04:48 & (0,--900) & 170 & $<$56 \\
S68N & 18:27:15.5 & 01:14:50 & (--900,0) & 310 & 50 \\
SMM6 & 18:27:25.4 & 01:12:02 & (--900,0) & 310 & 120 \\
HH108 & 18:33:07.6 & -00:35:48 & (--900,0) & 310 & 54 \\
CrA IRAS32 & 18:59:35.8 & -37:11:53 & (0,--900) & 130 & 148 \\
L673A & 19:18:04.6 & 11:14:12 & (--900,0) & 300 & $<$55 \\
L1152 & 20:35:19.4 & 67:42:30 & (--900,0) & 350 & 101 \\
L1172 & 21:01:44.2 & 67:42:24 & (--900,0) & 350 & 44 \\
L1251A & 22:34:22.0 & 75:01:32 & (--900,0) & 300 & $<$108 \\
L1251B & 22:37:40.8 & 74:55:50 & (--900,0) & 300 & $<$91 \\
IRAS23011 & 23:01:10.1 & 61:26:16 & (-600,0) & 730 & 57 \\
CB244 & 23:23:48.7 & 74:01:08 & (-900,0) & 250 & 56 \\
\enddata
\tablecomments{Class 0 sources have $T_{bol}$ $<$ 70 K and Class I
sources have 70 K $\le$ $T_{bol}$ $\le$ 650 K.  
All bolometric temperatures are from Mardones et al.\ 1997
except for the following sources which were calculated using the following
references: IRAS03282 (Barsony et al.\ 1998), L1535 (IRAS Point Source Catalog, 
Moriarty-Schieven et al.\ 1994), NGC2264G (Ward-Thompson et al. 1995), B228 
(IRAS Point Source Catalog, Reipurth et al. 1993), 
L1634 (IRAS Point Source Catalog, Cohen et al.\ 1985, Reipurth et al.\ 1993), 
MMS1, MMS4, MMS6, MMS9, MMS7, MMS8 (Chini et al.\ 1997), S68N (Hurt and Barsony
1996, Wolf-Chase et al.\ 1998), SMM6 (Hurt and Barsony 1996, Casali et 
al.\ 1993), L1152 (Myers et al.\ 1987, IRAS Point Source Catalog), L1172 (IRAS 
Point Source Catalog, Ladd et al.\ 1991), IRAS23011 (IRAS Point Source Catalog, 
Lefloch et al.\ 1996) and CB244 (IRAS Point Source 
Catalog, Launhardt et al.\ 1997).}
\end{deluxetable}

\clearpage
\begin{deluxetable}{llllll}
\tablewidth{0pt}
\tablecaption{List of Observed Lines}
\tablehead{\colhead{Molecule} & \colhead{Transition} & \colhead{Beamwidth} &
\colhead{$\eta_{mb}$} & \colhead{$\delta$v} & \colhead{Frequency} \\ & &
\colhead{(\arcsec)} & & \colhead{(\kms)} & \colhead{(GHz)}}
\startdata
H$^{13}$CO$^{+}$ & $J=3-2$ & 26 & 0.66 & 0.11 & 260.255478 \\
HCO$^{+}$ & $J=3-2$ & 26 & 0.66 & 0.11 & 267.557620
\enddata
\end{deluxetable}

\clearpage
\begin{deluxetable}{lllll}
\footnotesize
\tablewidth{0pt}
\tablecaption{Results}
\tablehead{\colhead{Source} & \colhead{Line} &
\colhead{$T_A^*$} & \colhead{Velocity} & \colhead{$\Delta$V} \\ & &
\colhead{(K)} & \colhead{(\kms)} & \colhead{(\kms)}}
\startdata
IRAS03235 & HCO$^{+}$ $J=3-2$ & 0.87$\pm$0.04 & 4.91$\pm$0.06 & 1.09$\pm$0.12 \\
-- & -- & 0.42$\pm$0.04 & 5.64$\pm$0.06 & -- \\
L1455 & HCO$^{+}$ $J=3-2$ & 2.65$\pm$0.10 & 4.29$\pm$0.06 & 1.33$\pm$0.12 \\
IRAS03256 & HCO$^{+}$ $J=3-2$ & 1.38$\pm$0.10 & 7.27$\pm$0.02 & 0.91$\pm$0.06 \\
NGC 1333 IRAS 2 & H$^{13}$CO$^{+}$ $J=3-2$ & 1.07$\pm$0.06 & 7.68$\pm$0.02 & 1.04$\pm$0.04 \\
-- & HCO$^{+}$ $J=3-2$ & 2.33$\pm$0.06 & 7.07$\pm$0.07 & 2.85$\pm$0.14 \\
-- & -- & 1.72$\pm$0.06 & 8.64$\pm$0.07 & -- \\
SSV13 & HCO$^{+}$ $J=3-2$ & 3.87$\pm$0.10 & 8.37$\pm$0.01 & 2.07$\pm$0.03 \\
IRAS03282 & H$^{13}$CO$^{+}$ $J=3-2$ & 0.21$\pm$0.04 & 7.13$\pm$0.04 & 0.65$\pm$0.12 \\
-- & HCO$^{+}$ $J=3-2$ & 2.12$\pm$0.03 & 6.79$\pm$0.08 & 0.95$\pm$0.16 \\
HH211 & HCO$^{+}$ $J=3-2$ & 2.41$\pm$0.06 & 9.24$\pm$0.08 & 1.16$\pm$0.15 \\
IRAS04166 & HCO$^{+}$ $J=3-2$ & 1.33$\pm$0.04 & 6.35$\pm$0.06 & 1.94$\pm$0.12 \\
-- & -- & 0.51$\pm$0.04 & 7.07$\pm$0.06 & -- \\
IRAS04169 & HCO$^{+}$ $J=3-2$ & 0.99$\pm$0.07 & 6.93$\pm$0.01 & 0.61$\pm$0.04 \\
L1551 IRS5 & HCO$^{+}$ $J=3-2$ & 6.11$\pm$0.10 & 6.58$\pm$0.01 & 1.14$\pm$0.01 \\
L1535 & HCO$^{+}$ $J=3-2$ & 1.32$\pm$0.06 & 5.43$\pm$0.01 & 0.78$\pm$0.02 \\
TMC-1A & HCO$^{+}$ $J=3-2$ & 1.27$\pm$0.04 & 5.97$\pm$0.06 & 1.45$\pm$0.12 \\
-- & -- & 1.04$\pm$0.04 & 6.58$\pm$0.06 & -- \\
L1634 & HCO$^{+}$ $J=3-2$ & 2.22$\pm$0.10 & 7.80$\pm$0.07 & 1.14$\pm$0.13 \\
-- & -- & 2.11$\pm$0.10 & 8.51$\pm$0.07 & -- \\
MMS1 & HCO$^{+}$ $J=3-2$ & 3.47$\pm$0.09 & 10.48$\pm$0.06 & 2.18$\pm$0.12 \\
-- & -- & 4.53$\pm$0.09 & 11.69$\pm$0.06 & -- \\
MMS4 & HCO$^{+}$ $J=3-2$ & 3.74$\pm$0.09 & 10.80$\pm$0.06 & 1.69$\pm$0.12 \\ 
-- & -- & 4.28$\pm$0.09 & 11.45$\pm$0.06 & -- \\
MMS6 & HCO$^{+}$ $J=3-2$ & 6.30$\pm$0.09 & 11.08$\pm$0.01 & 1.83$\pm$0.01 \\ 
MMS9 & HCO$^{+}$ $J=3-2$ & 4.99$\pm$0.10 & 11.49$\pm$0.01 & 1.70$\pm$0.02 \\ 
MMS7 & HCO$^{+}$ $J=3-2$ & 3.35$\pm$0.10 & 10.56$\pm$0.01 & 1.85$\pm$0.02 \\ 
MMS8 & HCO$^{+}$ $J=3-2$ & 5.05$\pm$0.11 & 11.49$\pm$0.01 & 1.18$\pm$0.02 \\ 
NGC2264G & H$^{13}$CO$^{+}$ $J=3-2$ & 0.18$\pm$0.04 & 4.56$\pm$0.06 & 0.74$\pm$0.12 \\
-- & HCO$^{+}$ $J=3-2$ & 1.24$\pm$0.03 & 4.57$\pm$0.01 & 1.28$\pm$0.03 \\ 
B228 & H$^{13}$CO$^{+}$ $J=3-2$ & 0.57$\pm$0.04 & 5.38$\pm$0.01 & 0.57$\pm$0.03 \\
-- & HCO$^{+}$ $J=3-2$ & 2.58$\pm$0.04 & 4.96$\pm$0.05 & 1.09$\pm$0.11 \\
-- & -- & 1.84$\pm$0.04 & 5.51$\pm$0.05 & -- \\
YLW 16 & HCO$^{+}$ $J=3-2$ & 0.90$\pm$0.10 & 4.51$\pm$0.05 & 2.18$\pm$0.11 \\
L146 & HCO$^{+}$ $J=3-2$ & 2.77$\pm$0.13 & 5.36$\pm$0.01 & 0.55$\pm$0.03 \\
S68N & H$^{13}$CO$^{+}$ $J=3-2$ & 0.81$\pm$0.10 & 8.83$\pm$0.03 & 1.30$\pm$0.08 \\
-- & HCO$^{+}$ $J=3-2$ & 2.35$\pm$0.09 & 7.22$\pm$0.06 & 4.35$\pm$0.12 \\
-- & -- & 3.87$\pm$0.09 & 9.57$\pm$0.06 & -- \\
SMM6 & HCO$^{+}$ $J=3-2$ & 4.51$\pm$0.11 & 1.91$\pm$0.03 & 7.69$\pm$0.01 \\
HH108 & HCO$^{+}$ $J=3-2$ & 1.13$\pm$0.03 & 10.38$\pm$0.06 & 1.36$\pm$0.12 \\
-- & -- & 1.75$\pm$0.03 & 11.00$\pm$0.06 & -- \\
CrA IRAS32 & HCO$^{+}$ $J=3-2$ & 1.32$\pm$0.04 & 5.08$\pm$0.06 & 1.61$\pm$0.12 \\
-- & -- & 2.27$\pm$0.04 & 5.95$\pm$0.06 & -- \\
L673A & HCO$^{+}$ $J=3-2$ & 2.46$\pm$0.09 & 6.92$\pm$0.01 & 1.05$\pm$0.02 \\
L1152 & HCO$^{+}$ $J=3-2$ & 2.43$\pm$0.11 & 2.45$\pm$0.01 & 0.68$\pm$0.02 \\
L1172 & H$^{13}$CO$^{+}$ $J=3-2$ & $<$0.05 & -- & -- \\
-- & HCO$^{+}$ $J=3-2$ & 1.08$\pm$0.03 & 2.54$\pm$0.06 & 1.12$\pm$0.12 \\
-- & -- & 1.14$\pm$0.03 & 3.16$\pm$0.06 & -- \\
L1251A & HCO$^{+}$ $J=3-2$ & 2.10$\pm$0.10 & -4.91$\pm$0.02 & 2.39$\pm$0.05 \\
L1251B & H$^{13}$CO$^{+}$ $J=3-2$ & 0.69$\pm$0.06 & -3.70$\pm$0.02 & 1.12$\pm$0.05 \\
-- & HCO$^{+}$ $J=3-2$ & 4.50$\pm$0.04 & -4.34$\pm$0.06 & 2.11$\pm$0.12 \\
-- & -- & 3.21$\pm$0.04 & -2.98$\pm$0.06 & -- \\
IRAS23011 & H$^{13}$CO$^{+}$ $J=3-2$ & 0.50$\pm$0.06 & -11.04$\pm$0.05 & 1.31$\pm$0.13 \\
-- & HCO$^{+}$ $J=3-2$ & 2.82$\pm$0.06 & -11.58$\pm$0.07 & 3.18$\pm$0.14 \\
-- & -- & 1.81$\pm$0.06 & -10.57$\pm$0.07 & -- \\
CB244 & HCO$^{+}$ $J=3-2$ & 2.73$\pm$0.04 & 3.75$\pm$0.06 & 1.61$\pm$0.12 \\
-- & -- & 1.80$\pm$0.04 & 4.62$\pm$0.06 & -- \\
\enddata
\end{deluxetable}

\clearpage
\begin{deluxetable}{lll}
\tablewidth{0pt}
\tablecaption{Line Asymmetries}
\tablehead{\colhead{Source} & \colhead{Asymmetry} & \colhead{Blue/Red}}
\startdata
IRAS03235 & -0.66$\pm$0.18 & 2.07 \\
L1455 &  -0.71$\pm$0.09 & - \\
IRAS03256 & -0.35$\pm$0.07 & - \\
NGC 1333 IRAS 2 & -0.59$\pm$0.11 & 1.35 \\
SSV13 & -0.11$\pm$0.02 & - \\
IRAS03282 & -0.52$\pm$0.17 & - \\
HH211 & 0.31$\pm$0.18 & -- \\
IRAS04166 & -0.97$\pm$0.18 & 2.61 \\
IRAS04169 & 0.07$\pm$0.05 & - \\
L1551 IRS 5 & 0.20$\pm$0.02 & - \\
L1535 & -0.60$\pm$0.04 & - \\
TMC-1A & -1.02$\pm$0.15 & 1.22 \\
L1634 & - & 1.06 \\
MMS1 & - & 0.77 \\
MMS4 & - & 0.87 \\
MMS6 & - & - \\
MMS9 & - & - \\
MMS7 & - & - \\
MMS8 & - & - \\
NGC2264G & 0.01$\pm$0.08 \\
B228 & -0.74$\pm$0.10 & 1.40 \\
YLW 16 & 0.45$\pm$0.07 & - \\
L146 & 0.40$\pm$0.14 & - \\
S68N & 0.57$\pm$0.07 & 0.74 \\
SMM6 & - & - \\
HH108 & 0.29$\pm$0.09 & 0.65 \\
CrA IRAS32 & 0.48$\pm$0.12 & 0.58 \\
L673A & 0.00$\pm$0.04 & - \\
L1152 & -0.63$\pm$0.06 & - \\
L1172 & 0.72$\pm$0.12 & 0.95 \\
L1251A & 0.16$\pm$0.05 & - \\
L1251B & -0.57$\pm$0.06 & 1.40 \\
IRAS23011 & -0.41$\pm$0.08 & 1.55 \\
CB244 & -1.02$\pm$0.14 & 1.52
\enddata
\end{deluxetable}

\clearpage
\begin{deluxetable}{llllll}
\tablewidth{0pt}
\tablecaption{Asymmetries -- Overall Sample}
\tablehead{\colhead{Class} & \colhead{N} &
\colhead{Blue} & \colhead{Neither} & \colhead{Red} & \colhead{Blue Excess}}
\startdata
0 & 26 & 15 & 4 & 7 & 0.31\\
I & 16 & 8 & 5 & 3 & 0.31 \\
Overall & 42 & 23 & 9 & 10 & 0.31 \\
\enddata
\end{deluxetable}

\clearpage
\begin{deluxetable}{lllll}
\footnotesize
\tablewidth{0pt}
\tablecaption{Asymmetries by source}
\tablehead{\colhead{Source} & \colhead{CS} & \colhead{H$_{2}$CO} & 
\colhead{HCO$^{+}$} & \colhead{HCN}}
\startdata
NGC 1333 IRAS 4A & B & B & B & B \\
L1157 & B & B & B & B \\
L1251B & B & B & B & B \\
NGC 1333 IRAS 4B & B & B & B & - \\
IRAS16293 & B & B & B & - \\
L483 & B & B & R & B \\
B335 & B & B & B & R \\
\\
IRAS03256 & B & N & B & - \\
L1527 & N & B & B & - \\
Serpens SMM4  & N & B & B & N \\
CB244 & B & N & B & N \\
\\
L483 & B & B & R & B \\
VLA1623 & B & B & R & N \\
S68N & B & B & R & - \\ 
\\
IRAS03235 & N & N & B & - \\ 
IRAS03282 & N & N & B & - \\
IRAS04166 & N & N & B & - \\
L1535 & N & N & B & - \\
L1152 & N & N & B & N \\
TMC-1A & N & N & B & - \\
L1455 & - & N & B & - \\
B228 & - & N & B & - \\
\\
NGC 1333 IRAS 2 & N & R & B & N \\
L1251A & R & B & N & N \\
\\
L1172 & - & N & R & B \\
\\
L1448-N & N & N & N & N \\
SSV13 & N & N & N & - \\
IRAS04169 & N & N & N & - \\
L723 & N & N & N & - \\
L673A & N & N & N & N \\
\\
HH211 & N & N & R & - \\
L1551 & R & N & N & - \\
L146 & N & N & R & R \\
HH108 & N & N & R & N \\
CrA IRAS32 & N & N & R & - \\
L146 & N & N & R & R \\
L1172 & - & N & R & B \\
\\
L1448-C & B & R & R & B \\
L1551-NE & R & R & B & - \\
Serpens SMM1 & B & R & R & - 
\enddata
\end{deluxetable}

\clearpage
\begin{figure}
\caption{HCO$^{+}$ and H$^{13}$CO$^{+}$ $J=3-2$ spectra toward the center
of five Class 1 sources and four Class 0 sources.  
These sources all show a blue asymmetry in the
HCO$^{+}$ $J=3-2$ line characteristic of collapse.  The solid line is the
HCO$^{+}$ spectrum and the dashed line is the H$^{13}$CO$^{+}$ spectrum.
The dashed vertical line is the velocity of the N$_{2}$H$^{+}$ $J=1-0$ line
observed by Mardones et al.
The intensity and velocity scales for the L1251B spectra are 
shown on the right and top side of its panel, respectively.  The velocity scale
for the IRAS23011 spectra is the same as that of L1251B.}
\end{figure}

\begin{figure} 
\caption{HCO$^{+}$ and H$^{13}$CO$^{+}$ $J=3-2$ spectra toward the center
of one Class 1 and five Class 0 sources that all show a red asymmetry in the
HCO$^{+}$ $J=3-2$ line. The solid line is the
HCO$^{+}$ spectrum and the dashed line is the H$^{13}$CO$^{+}$ spectrum.  
The dashed vertical line is the velocity of the N$_{2}$H$^{+}$ $J=1-0$ line 
observed by Mardones et al.}
\end{figure}

\begin{figure}
\caption{HCO$^{+}$ $J=3-2$ spectra toward the center of three Class I 
sources and seven Class 0 sources that do not show any clear asymmetry in 
the HCO$^{+}$ $J=3-2$ line.  The solid line is the
HCO$^{+}$ spectrum and the dashed line is the H$^{13}$CO$^{+}$ spectrum.  
The dashed vertical line is the velocity 
of the N$_{2}$H$^{+}$ $J=1-0$ line observed by Mardones et al.}
\end{figure}

\begin{figure} 
\caption{HCO$^{+}$ $J=3-2$ spectra toward the center of six Class I and three
Class 0
sources that do not show any clear asymmetry in the HCO$^{+}$ $J=3-2$ line.
The dashed vertical line is the velocity of the N$_{2}$H$^{+}$ $J=1-0$ line 
observed by Mardones et al.
The velocity scale for the L1251A spectra is shown on the 
top side of its panel.}
\end{figure}

\begin{figure} 
\caption{Histogram of the velocity of the H$^{13}$CO$^{+}$ $J=3-2$ line minus that of  the N$_{2}$H$^{+}$ $J=101-012$ line.}
\end{figure}

\begin{figure} 
\caption{The $\delta$v of the observed line profiles vs. bolometric temperature.}
\end{figure}

\begin{figure} 
\caption{The top panel is the $\delta$v of the HCO$^{+}$ $J=3-2$ line
profile vs. that of the CS $J=2-1$ line.  The bottom
panel is the $\delta$v of the HCO$^{+}$ $J=3-2$ line
profile vs. that of the H$_{2}$CO $J=2_{12}-1_{11}$ line.}
\end{figure}

\begin{figure} 
\caption{The top row is a set of six models following the temporal
evolution of the HCO$^{+}$ $J=3-2$ line profile.  The middle row
follows the evolution of the H$_{2}$CO $J=2_{12}-1_{11}$ line profile.
The bottom row follows the evolution of the CS $J=2-1$ line profile.
The intensity scale is in T$_{A}^{*}$.  The dashed horizontal lines are
the blue-red ratios.}
\end{figure}

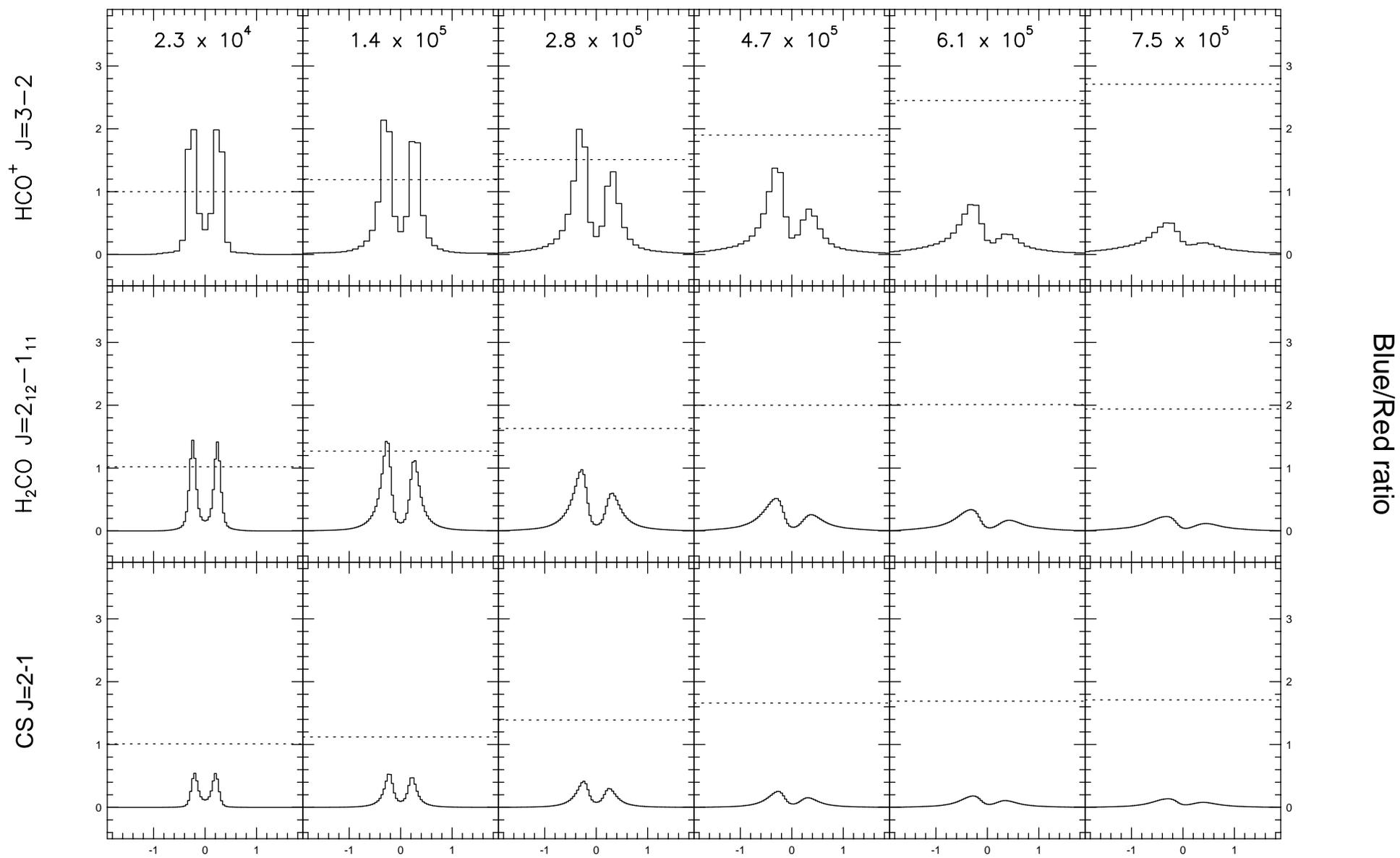
\end{document}